\documentclass[preprint,showpacs,preprintnumbers,amsmath,amssymb]{revtex4}

\input epsf

\begin{document}

\title{Proper orthogonal decomposition of solar photospheric motions}

\author{A. Vecchio$^1$, V. Carbone$^1$, F. Lepreti$^1$, L. Primavera$^1$,
L. Sorriso-Valvo$^{2,1}$, P. Veltri$^1$, G. Alfonsi$^3$, Th. Straus$^4$}

\affiliation{
$^1$ Dipartimento di Fisica, Universit\`a della Calabria,
and Istituto Nazionale di Fisica della Materia, Unit\`a di Cosenza,
87030 Rende (CS), Italy \\
$^2$ Laboratorio Regionale LYCRYL - INFM/CNR, 87030 Rende (CS), Italy\\
$^3$ Dipartimento di Difesa del Suolo, Universit\`a della Calabria, 87030 Rende (CS), Italy\\
$^4$ INAF - Osservatorio Astronomico di Capodimonte, Napoli, Italy
}

\begin{abstract} 
The spatio-temporal dynamics of the solar photosphere is studied by performing
a Proper Orthogonal Decomposition (POD) of line of sight velocity fields computed
from high resolution data coming from the MDI/SOHO instrument. 
Using this technique, we are able to identify and characterize the different
dynamical regimes acting in the system. Low frequency oscillations, with frequencies
in the range $20$--$130$~$\mu$Hz, dominate the most energetic POD modes 
(excluding solar rotation), and are characterized by spatial patterns with typical 
scales of about $3$~Mm. Patterns with larger typical scales of  $\simeq 10$~Mm, 
are associated to p-modes oscillations at frequencies of about $3000$~$\mu$Hz. 

\end{abstract}

\pacs{96.60.Mz; 92.60.Ek; 96.60.Ly; 89.75.-k}

\date{\today}

\maketitle


The solar photosphere is an interesting example of a system exhibiting
complex spatio-temporal behavior, a quite common feature in a wide range
of systems far from equilibrium. The early research in pattern
formation focused on the presence of simple periodic structures, while
the main questions currently addressed concern regimes characterized by higher
complexity, that is, patterns that are more irregular in space and time. This is
often related to the occurrence of intermediate states between order and
turbulence~\cite{pattern}.
High resolution images of the solar photosphere show that the whole
surface of the Sun is covered by a cellular pattern,
the so called {\em solar granulation}, consisting in bright cells (granules),
with roughly $10^3$~km size, surrounded by a dark background of intergranular
lanes cf.~\cite{granulation,Stix}.
This structure represents the manifestation of the complex convective
dynamics, at high Rayleigh numbers, occurring in the subphotospheric 
layers~\cite{Stix,simulazioni}. 
Other patterns are also observed in the photosphere,
namely the supergranulation ($\simeq 3\times 10^4$~km size)~\cite{helio-p,rieutord} 
and giant cells of size about $10^5$~Km, perhaps the result of global convection over 
the whole solar convective depth~\cite{giant}.
In the solar photosphere the convective dynamics coexist with the
contribution of solar oscillations.
The interior of the Sun, because of the gravitational field and density
stratification, behaves like a resonant cavity and supports
the excitation of global oscillations~\cite{helio-p,helio-g,helio}.
Depending on the restoring force, two main types of waves
can be excited, namely acoustic high frequencies p-modes~\cite{helio-p,helio}
(in the range $1000$--$5000$~$\mu$Hz~\cite{ulrich}) and gravitational
low frequency g-modes~\cite{helio-g,helio} (in the range $1$--$200$~$\mu$Hz).
The Fourier $k$--$\omega$ spectra of line of sight photosperic velocity
fields show ridges corresponding to the discrete p-modes~\cite{deubner} (see fig.\ref{kw}).
On the other hand, discrete frequencies in the low-frequency 
range (g-modes) are not recognized,
but rather the power observed in this range is spread over a continuum,
commonly attributed to the solar turbulent convection.
Indications for the presence of $160$~min discrete oscillations within
a continuum have been reported~\cite{helio-g}, followed by contradictory
evidences~\cite{against}. Despite these attempts, no unambiguous mode
identification and classification has been established so far.


In the present paper, we use the Proper Orthogonal Decomposition (POD)
as a powerful tool to investigate the dynamics of stochastic spatio-temporal
fields.
In astrophysical contexts, POD has been recently used to analyze the
spatio-temporal dynamics of the solar cycle~\cite{ciclo}.
Introduced in the context of turbulence~\cite{POD}, the POD decomposes a
field $u({\bf r},t)$ as
$
u({\bf{r}},t) = \sum_{j=0}^{\infty} a_j(t) \Psi_j(\bf{r})
$~(1),
the eigenfunctions $\Psi_{j}$ being constructed by maximizing the
average projection of the field onto $\Psi_{j}$, constrained to the unitary
norm. Averaging leads to an optimization problem that can be cast as
$
\int_{\Omega}\langle u{(\bf{r},t)},{u(\bf{r'},t)}\rangle
\Psi({\bf{r'}}){\bf{dr'}}=\lambda \Psi(\bf{r})
$~(2),
where $\Omega$ represents the spatial domain and brackets represent time averages.
The integral equation~(2) provides the eigenfunctions~$\Psi_{j}$ and
a countable, infinite set of ordered eigenvalues $\lambda_j \geq
\lambda_{j+1}$, each representing the kinetic energy of the $j$-th~mode.
Thus, POD builds up the basis functions, which are not given 
\textit{a priori}, but rather obtained from observations.
The time coefficients~$a_j(t)$ are then computed from the projection of
the data on the corresponding basis functions $\Psi_j(\bf{r})$ so that the sum~(1), when
truncated to $N$~terms, contains the largest possible energy with respect to any other linear
decomposition of the same truncation order.
The POD basis functions are optimal in turbulence with respect to the
classical Fourier analysis, where the basis functions are not proper eigenfunctions of the signal.
This method is particularly appropriate when analyzing complex physical systems, 
where different dynamical regimes coexist. POD allows
to identify these regimes and to characterize their energetics and their spatial structure.
We show that POD is able to capture the main aspects of the spatio-temporal dynamics 
of the solar photosphere. 


The line of sight velocity fields $u(x,y,t)$ ($x,y$ being the
coordinates on the surface of the Sun) used in this work have been obtained from images
acquired by the Michelson Doppler Imager (MDI) instrument mounted on the SOHO
spacecraft~\cite{soho}.
The image size is $695 \times 695$ pixels, with a spatial resolution of
about $0.6$~arcsec ($1$~arcsec $\simeq 725$~km on the surface). The time
series spans a time interval $T = 886$~min, images being sampled every minute. 
The top panel of Figure~\ref{fig_pod_eigenf} shows one of the snapshots of the dataset,
together with its spatial spectrum. These data have been already studied in the past with Fourier
techniques~\cite{fourier-mdi}, and the $k$--$\omega$ power spectrum of the velocity 
signal is shown in figure~\ref{kw}.


The POD of the velocity field~$u(x,y,t)$  yields a set of eigenfunctions
$\Psi_{j}(x,y)$ and coefficients~$a_{j}(t)$, as well as the sequence of
eigenvalues $\lambda_{j}$ ($j = 0,1,\dots,886$, sorted in decreasing
energetic content).
%
%
Most of the energy is associated with the first POD mode, accounting
for the line of sight component of solar rotation. As this is neither the expression 
of solar turbulence nor oscillations, this first POD mode will be ignored throughout this paper.
The rest of the energy is decreasingly shared by the following
885 modes, so that, for example, only $4\%$ is associated to the second ($j=1$) POD mode.
In laboratory turbulent flows, POD attribute almost $90\%$ of total
energy to the typical large-scale coherent structures, confined to $j \leq 2$~\cite{turbo}.
In our case, $90\%$ of energy (excluding the rotation) is contained in
the first $140$ modes.
This indicates the absence of dominating large-scale structures and the
presence of some turbulent dynamics related to nonlinear interactions among different
modes and structures at all scales.


We turn now the attention to the eigenfunctions, namely the spatial
patterns, found by POD, and reported in figure~\ref{fig_pod_eigenf} for three modes. The power
spectra $|\Psi_{j}(k)^{2}|$ obtained from Fourier transform of the eigenfunctions $\Psi_{j}(x,y)$
are also reported in figure~\ref{fig_pod_eigenf}, for the same three modes, as a function of
the wavevector $k=(k_x^2+k_y^2)^{1/2}$.
It can be observed that the patterns are rather disordered, with a broad spectrum.
Moreover, modes can be separated in three different groups, accordingly
to grain size, as is shown in figure~\ref{fig_pod_eigenf}.
The $1\leq j\leq 11$, most energetic modes display a pattern with very
fine structures, recalling the photospheric granulation, with a broad spectrum indicating the
complex nature of the POD modes, and show a center-limb modulation.
Conversely, eigenfunctions of the modes $j=12$ and $j=13$
present a coarser pattern, and their (still broad) spectra show a number of ridges.
The further lower energy modes cannot be precisely classified, and seem to
present a mixture of the previous characteristics, both in grain size
and in spectral shape.
The spatial pattern associated to each mode, although complicated, can
be quantitatively characterized by computing the integral scale length
$
L_{j}={\int_{0}^{\infty}\left|\psi_{j}(k)\right|^{2}k^{-1}dk}/{\int_{0}^{\infty}\left|\psi_{j}(k)\right|^{2}dk}
$ 
which represents the energy containing scale of classical turbulence~\cite{pope}.
This allows to estimate the typical scale for the first ten fine grained
modes $L\simeq 3$~Mm, while for the coarse grained modes $L\simeq 10$~Mm.
   \begin{figure}[ht]
   \caption{
   Left column, from top to bottom: one snapshot
   of the original data, and the POD eigenfunctions
   $\Psi_{j}(x,y)$ in the plane $(x,y)$ for the three modes $j=8$, $j=12$ and $j=49$.
   Each image covers about $300 \times 300$~Mm$^{2}$ on the solar surface.
   Right column: the wave vector spectra $|\Psi_{j}(k)|^2$
   of the corresponding images. On the second and third
   spectra, the wave vector spectra obtained from the $k$--$\omega$ cut,
   respectively at $\omega=76$~$\mu$Hz and $\omega=3254$~$\mu$Hz,
   are superimposed (dotted lines, in arbitrary units).}
   \label{fig_pod_eigenf}
   \end{figure}
  \begin{figure}[ht]
  \caption{The $k$--$\omega$ spectrum of the dataset.}
  \label{kw}
  \end{figure}


The eigenfunctions described above are associated with the corresponding
coefficients, accounting for the temporal evolution of each mode.
In figure~\ref{fig_pod_coeff} we report, for the same modes represented
in figure~\ref{fig_pod_eigenf}, the time behaviour of the coefficients
$a_{j}(t)$, together with their Fourier spectra $|a_{j}(\nu)|^2$.
As can be seen, for all the eigenfunctions the spectra of the time
coefficients present two peaks, located in two well defined and separated frequency ranges.
The fine grained eigenfunctions ({\it e. g.} $j=8$) are associated with time coefficients
dominated by low-frequency oscillations. For the coarse grained patterns ({\it e. g.} $j=12$) the
high frequency peak becomes of the same order, and even slightly higher than the low frequency one. 
For the intermediate cases ({\it e. g.} $j=49$), the amplitudes of the
two peaks are again of the same order, but with weak prevalence of the low frequencies. 
For each POD mode, the high frequency peaks can
be identified from the spectra, lying in the range $3250$--$3550$~$\mu$Hz.
In a similar way, for the low frequency oscillations, we find frequencies in the
range $20$--$127$~$\mu$Hz.
   \begin{figure}[ht]
   \caption{Left column: time evolution of POD
   coefficients $a_{j}(t)$ ($m/s$) for the same three modes as in figure~\ref{fig_pod_eigenf}.
   Right column: the corresponding frequency spectra $|a_{j}(\nu)|^2$ ($m^2/s^2$).}
   \label{fig_pod_coeff}
   \end{figure}


We try now to understand the results observed with POD.
Let us first focus on the high-frequency modes.
Both the informations on the involved spatial scales and the measured
frequencies indicate that solar p-mode contributions are identified by POD. 
Indeed, the frequencies we obtain (for instance, $\nu=3405$~$\mu$Hz for $j=12$, $\nu=3367$~$\mu$Hz 
for $j=13$) are compatible with the p-modes observed using Fourier techniques~\cite{deubner,modiP,Stix}, 
and have been predicted by helioseismological models cf.~\cite{helio}. 
The spatial pattern of about $10$~Mm, which POD associates with high
frequency modes, is in agreement with the horizontal coherence length attributed to solar
p-modes. In order to confirm this, we compared the $k$--$\omega$ classical
results with the POD spectra. Choosing, for example, the POD mode $j=12$, we performed a
cut of the $k$--$\omega$ spectrum at the measured peak frequency ($3254$~$\mu$Hz).
We thus obtain a wavelength spectrum, reported on figure~\ref{fig_pod_eigenf}, that can
be compared with the $j=12$ eigenfunction spectrum. As can be seen, the spikes observed in
the POD spectrum qualitatively correspond to the typical ridges structure of the
$k$--$\omega$ spectrum.


Concerning the low-frequency dominated modes, it is interesting to note
that the measured frequency range, mentioned above, is compatible with the theoretical
results on solar gravitational modes~\cite{Stix,helio}. However, because of the
resolution limit, related to the finite length of the time series, the low frequency oscillations
detected by POD can not be unambiguosly attributed to discrete modes, rather than to a continuum.
The evidence of center-limb modulation shows that such modes are mainly
associated with the contribution of horizontal velocities. To confirm this, we tried to
apply the POD technique to a reduced square of about $100 \times 100$ pixels taken at the center
of each image, where horizontal components of the velocity are weaker. In this case,
the low-frequency oscillations are no longer observed (see figure~\ref{fig_pod_test}).
As for the p-modes, the $k$--$\omega$ cut can be performed at the peak
frequency of, for example, mode $j=8$ ($76$~$\mu$Hz). The comparison with the POD
eigenfunction wavelength spectrum is shown in figure~\ref{fig_pod_eigenf}, and again
reveals a qualitative agreement between the two techniques. The characteristic
integral scale computed for the $k$--$\omega$ cut is also in agreement with the
corresponding POD integral scale. Since the POD basis functions are not chosen a priori,
the association between small-scale eigenfunctions and low-frequencies is reliable.
In order to test this, we reduced the resolution of the data by
performing a spatial $50\times 50$ pixels running average of the fields before applying POD.
In this case, the low-frequency peaks in the coefficient spectra are completely lost
(see figure~\ref{fig_pod_test}).
  \begin{figure}[ht]
  \caption{Upper panels: the POD eigenfunction $\Psi_6(x,y)$ 
  (left, the center-limb modulation is evident) and the
  Fourier spectra (right) obtained from two reduced square boxes of size $100 \times 100$ pixels,
  taken at the center (dotted) and at the border (full) of the image (see left panel).
  Bottom panels: $50\times 50$ pixels running averaged data at a given time
  (left), and the $j=9$ POD mode wave vector spectrum (right) for both the smoothed field
  (full) and the original field (dotted).}
  \label{fig_pod_test}
  \end{figure}


In conclusion, we presented the first application of POD on high
resolution solar photospheric velocity fields, which represents an example of 
convective turbulence in high Rayleigh numbers natural fluids. 
POD is able to capture the main energetic and spatial features of the solar photosphere.
In particular, the dynamical processes of interest are automatically separated from the most energetic,
but uninteresting mode witch captures solar rotation. Two main oscillatory processes,
well separated in frequency, are detected in all the other modes: high frequency oscillations
in the range $3250$--$3550$~$\mu$Hz and low frequency oscillations in the range
$20$--$130$~$\mu$Hz.
The high frequency waves are the well known acoustic p-modes and their properties, as
obtained from POD, are in agreement with previous results based on Fourier techniques.
On the other hand, low frequency oscillations, with frequencies compatible with those expected
from the theory of solar g-modes, prevail in the most energetic POD modes
(excluding solar rotation), which are characterized by spatial eigenfunctions with typical 
scales of $\simeq 3$~Mm. The clear association between low frequency oscillations and small spatial
scales, which are close to the solar granulation, is the most interesting and most surprising
result provided by our analysis.
This suggests the presence of a strong coupling between 
low frequency modes and the turbulent convection, which could have implications on 
the inversion of time-distance measurements of sound waves~\cite{duvall}, 
also used to infer the properties of the turbulent convection below the surface. 
Such coupling could be related to the dishomogeneities and small scale structures 
arising from the highly nonlinear dynamics of the convective layers. 
This point needs to be investigated in future theoretical studies. 
A further improvement of this analysis can be expected 
from its application to longer time series.

{\bf Acknowledgments}
We acknowledge useful discussions with A. Mangeney.
Th.S. was supported by MIUR.


\end{document}